\documentclass[prb,floatfix,twocolumn,superscriptaddress,showpacs,amsmath,amssymb,showpacs]{revtex4-2}

\usepackage{graphicx}
\usepackage{epstopdf}
\usepackage{epsfig}
\usepackage{epsf}
\usepackage{url}
\usepackage[USenglish]{babel}
\usepackage{hyperref}
\allowdisplaybreaks
\usepackage{verbatim}
\usepackage{natbib}
\usepackage{amsmath, nccmath}
\usepackage[export]{adjustbox}
\usepackage{dcolumn}
\usepackage{bm}
\usepackage{bbm}
\usepackage{lipsum}

\usepackage{xcolor}
\newcommand{\chem}[1]{\ensuremath{\mathrm{#1}}}

\newcommand{\etal}{\textit{ et al.}}

\usepackage[section]{placeins}

\begin{document}

	\title{ Photo-induced insulator-metal transition in paramagnetic (V$_{1-x}$Cr$_{x}$)$_2$O$_3$}

\author{Jiyu Chen}

\affiliation{Department of Physics, University of Fribourg, 1700 Fribourg, Switzerland}
\author{Francesco Petocchi}
\affiliation{Department of Physics, University of Geneva, 1211  Geneva, Switzerland}
\author{Viktor Christiansson}
\affiliation{Department of Physics, University of Fribourg, 1700 Fribourg, Switzerland}
\author{Philipp Werner}
\affiliation{Department of Physics, University of Fribourg, 1700 Fribourg, Switzerland}
	\begin{abstract}
Pump-probe experiments with femtosecond time resolution allow to disentangle the electronic dynamics from the lattice response and thus provide valuable insights into the non-equilibrium behavior of correlated materials. In Cr-doped V$_2$O$_3$, a multi-orbital Mott-Hubbard material which has been intensively investigated for decades, time-resolved experiments reported a photo-induced insulator-metal transition leading to a transient metal state with nonthermal properties. Here, we combine non-equilibrium dynamical mean-field theory with realistic first principles modeling to simulate the ultrafast response of this material to a laser excitation. Our calculations reproduce the insulating initial state, with orbital occupations in agreement with experiment, and reveal an ultrafast pump-induced gap filling associated with a charge reshuffling between the $e_g^\pi$ and $a_{1g}$ orbitals. However, in contrast to the related compound VO$_2$, the electronic system thermalizes within a few tens of femtoseconds and we find no evidence for the existence of a metastable nonthermal metal. This suggests that the reported nonthermal behavior in the experiments may be associated with the mismatch between the electronic and lattice temperatures. 
\end{abstract}

\maketitle

\section{Introduction}

Several recent experiments revealed nonthermal metal states in photo-excited insulators, which persist for hundreds of femto-seconds~(fs) or even pico-seconds~(ps) despite a complete gap filling. Examples are 1$T$-TaS$_2$ \cite{ligges2018,dong2023,petocchi2022,petocchi2023}, where this behavior has been associated with the excitation of spin-triplet states in strongly coupled bilayers, and VO$_2$ \cite{morrison2014,wegkamp2014,chen2023}, where a theoretical analysis suggested an orbital reshuffling of charge as the origin of the nonthermal properties of the photo-induced metal. A third example is Cr-doped V$_2$O$_3$, where pump-probe experiments also report evidence for a photo-induced metal with long-lived nonthermal electron distributions \cite{lantz2017,babich2020}.

Vanadium sesquioxide V$_2$O$_3$ has been widely studied as a prototype multi-orbital Mott-Hubbard system \cite{held2001,keller2004,laad2006,poteryaev2007,toschi2010,rodolakis2010,hansmann2013}. At ambient pressure, V$_2$O$_3$ undergoes a metal-insulator transition (MIT) at $T_c\approx$~160~K, associated with a first-order structural transition. Above $T_c$, the corundum structure hosts a paramagnetic metal (PM) phase, while below $T_c$, the system exhibits a monoclinic antiferromagnetic insulator (AFI) phase. 
A different paramagnetic insulator (PI) phase is found at $T> T_c$ by chromium substitution of vanadium \cite{mcwhan1969, mcwhan1970, mcwhan1973}. This doping-induced first-order metal-to-insulator phase transition from the PM to PI phase is associated with a small discontinuity of the lattice parameter, without a change of lattice symmetry~\cite{dernier1970,robinson1975}. 
The Cr substitution has often been regarded as equivalent to a negative pressure. However, more recent studies found microscale phase separation in Cr-doped V$_2$O$_3$ at ambient pressure and interpreted the Cr atoms as ``condensation nuclei" in a percolative PM-PI phase transition, challenging the negative pressure interpretation \cite{lupi2010,hansmann2013}. 

\begin{figure}[t]
	\includegraphics[width=0.75\linewidth]{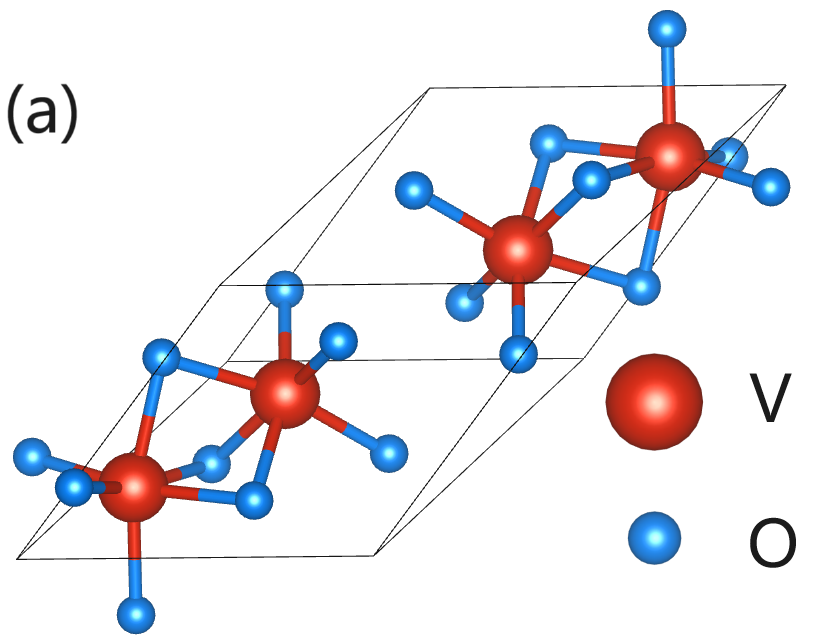}
		\centering
	\includegraphics[width=0.97\linewidth]{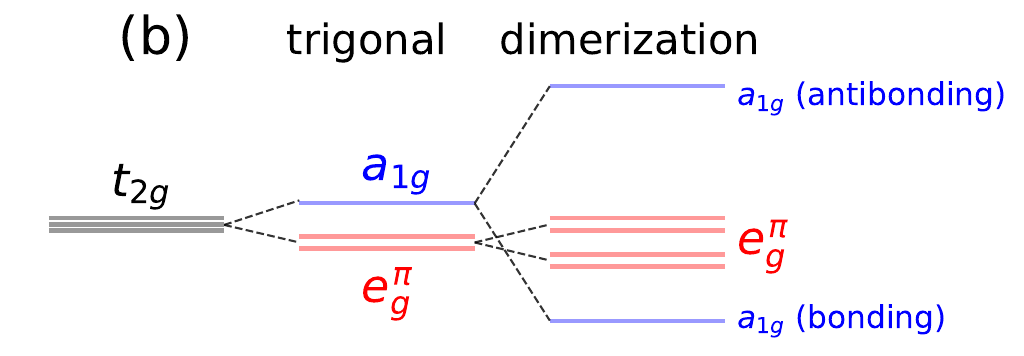}
	\includegraphics[width=\linewidth]{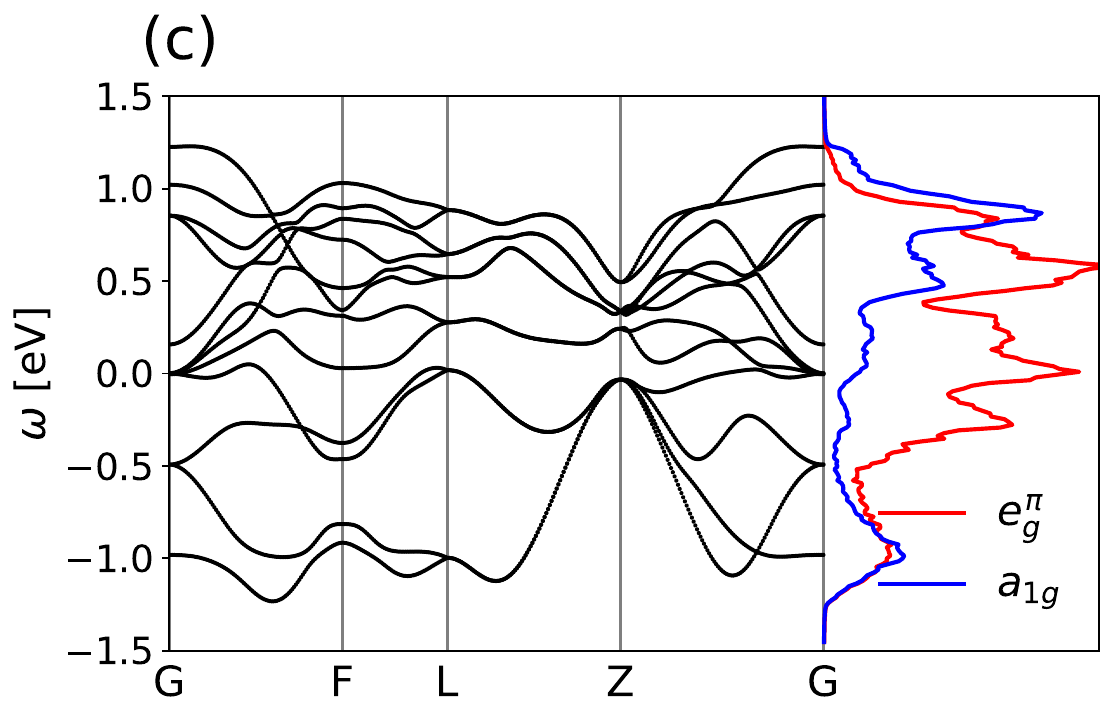}
	\caption{(a) Lattice structure of V$_2$O$_3$, with the four-atom unit cell marked in black. Red spheres: V atoms, blue spheres: O atoms.  (b) Energy level scheme of a non-interacting V-V dimer, including the effects of the trigonal crystal field splitting and the V-V dimerization. The totally six orbitals of each dimer host four electrons.
	(c) Wannier interpolated bandstructure (black curves) for PI (V$_{1-x}$Cr$_{x}$)$_2$O$_3$ ($x=0.038$). The orbital-projected densities of states are plotted on the right side. 
	}
	\label{fig:bands}
\end{figure}

The high-temperature paramagnetic V$_2$O$_3$ system exhibits a corundum crystal structure in which the V$^{3+}$ cations are surrounded by oxygen octahedra. This leads to a 3$d^{2}$ electronic configuration of V$^{3+}$, with empty high energy $e_g$ orbitals and two electrons in the three low energy $t_{2g}$ orbitals. A trigonal distortion of the crystal field further lifts the three-fold degeneracy of the $t_{2g}$ orbitals, resulting in one non-degenerate $a_{1g}$ orbital oriented along the $c$ axis and two degenerate $e_g^\pi$ orbitals oriented predominantly in the hexagonal plane. Furthermore, the vanadium atoms are dimerized by sharing an octahedral surface. This dimerization decreases the energy of the bonding $a_{1g}$ orbital, which becomes partially occupied in the ground (see Fig.~\ref{fig:bands}). 

Polarized X-ray spectroscopy experiments demonstrated that the two valence electrons of V$^{3+}$ form a high spin $S=1$ state, with an admixture of $e^\pi_ga_{1g}$ and $e^\pi_ge^\pi_g$ configurations \cite{park2000}, suggesting a multi-orbital interpretation of the material. Several theoretical studies \cite{held2001,keller2004,laad2006,poteryaev2007,toschi2010,rodolakis2010,hansmann2013} combining {\it ab-initio} calculations in the local density approximation (LDA) \cite{kohnsham1965} with dynamical mean-field theory (DMFT) \cite{georges1996} considered a three-orbital Kanamori-Hubbard description of the system with ($x=0.038$, PI) and without ($x=0$, PM) Cr doping. The LDA calculations~\cite{saha-dasgupta2009} predict that the Cr doping of V$_2$O$_3$ narrows the $t_{2g}$ bandwidth $W$ and thus increases the $U/W$ ratio, which underlies the PM to PI phase transition.  The imbalance in the occupation of the $a_{1g}$ and $e_{g}$ orbitals for both the PI and PM is due to the trigonal splitting of the $a_{1g}$ and $e_{g}$ orbitals.  The large Hund's coupling $J$ used in these previous studies favors the high spin state, in agreement with the experiment by Park {\it et al.}~\cite{park2000}.

Recently, the nonequilibrium properties of (V$_{1-x}$Cr$_{x}$)$_2$O$_3$ have been investigated with multiple time-resolved tools \cite{mansart2010,lantz2017,babich2020}. By combining time-resolved photoelectron spectroscopy, reflectivity and X-ray diffraction experiments, Lantz\etal~observed an instantaneous Mott gap collapse in PI (V$_{1-x}$Cr$_{x}$)$_2$O$_3$ ($x=0.038$) after a 35 fs laser pulse with energy 1.55 eV \cite{lantz2017}. They reported the existence of a nonthermal transient metal phase lasting for hundreds of fs and characterized by the overpopulation of the $a_{1g}$ orbital.

Here, we study the photo-induced dynamics in a realistic model of (V$_{1-x}$Cr$_{x}$)$_2$O$_3$ ($x=0.038$) using nonequilibrium DMFT~\cite{aoki2014}. Our three-orbital calculations with {\it ab initio} derived hopping parameters and realistic interactions yield an equilibrium Mott insulating phase with small gap, as observed in the experiments \cite{lupi2010,babich2020}. Nonequilibrium simulations yield a rapid metalization after a photo-doping pulse, but also a fast thermalization of the electronic system within a few tens of fs. We will show that both the photo-excitation and the heating result in a significant charge transfer from the $e^{\pi}_{g}$ to the $a_{1g}$ orbitals.

This paper is structured as follows. Section~\ref{sec:method} describes the {\it ab initio} modeling  and the DMFT simulations of the pulse-excited system, Sec.~\ref{sec:results} presents the time-resolved orbital occupations and spectra, while Sec.~\ref{sec:conclusions} provides a brief discussion and conclusion.

\section{Model and Method} 
\label{sec:method}

To derive a realistic model for paramagnetic insulating (V$_{1-x}$Cr$_{x}$)$_2$O$_3$, we start from the experimental lattice structure for $x = 0.038$ from Ref.~\onlinecite{dernier1970}, perform density functional theory (DFT) calculations using \textsc{Quantum} ESPRESSO \cite{giannozzi2017}, and downfold to the $t_{2g}$ orbitals using Wannier90 \cite{pizzi2020}. 
Figure~\ref{fig:bands}(c) shows the Wannier interpolated DFT band structure in the energy range $-1.5~\text{eV} \le \omega \le 1.5~\text{eV}$. The $e^{\pi}_{g}$ and  $a_{1g}$ bands strongly  overlap in this energy range, but the local density of states (DOS) of the $a_{1g}$ orbital is $\sim 0.13$~eV higher than for the $e^{\pi}_{g}$ orbitals, due to the trigonal crystal field splitting. The $a_{1g}$ DOS exhibits a strong bonding-anti-bonding feature. 
At the DFT level, the orbital fillings per spin are ($a_{1g}$ : $e_{g1}^\pi$ : $e_{g2}^\pi$) = (0.56 : 0.72 : 0.72).

The time-dependent low-energy Hamiltonian reads
\begin{equation}
	\begin{aligned}\label{eq:hamiltonian}
		\hat{\mathcal{H}}(t) =& \sum_{\mathbf{R},\mathbf{R'}}\sum_{i}\Big\{\sum_{j} \sum_{\alpha\beta,\sigma} h^{ij}_{\alpha\beta}(\mathbf{R}-\mathbf{R'},t)d_{\mathbf{R}\alpha \sigma}^{i\dagger} d^{j}_{\mathbf{R'}\beta \sigma}\nonumber\\
		&\hspace{14mm}-\sum_{\alpha \sigma}\mu n^{ai}_{\mathbf{R}\alpha\sigma}+\mathcal{H}_{\text{K}}^{i}(\mathbf{R})\Big\},
	\end{aligned}
\end{equation}
where  $\mathbf{R}$ labels the unit cell, $i, j\in \{1,2,3,4\}$ label the V atoms in a given unit cell, $\alpha,\beta \in \{1,2,3\}$ label the three $t_{2g}$ orbitals and $\sigma=\{\uparrow,\downarrow\}$ denotes spin. $n_{\alpha\sigma}$ is the occupation per spin of orbital $\alpha$ and $\mu$ the chemical potential. 
The hopping parameters $h^{ij}_{\alpha\beta}(\mathbf{R},t=0)$ are extracted from the first principles calculation. The corresponding bandstructure and DOS are plotted in Fig.~\ref{fig:bands}(c). 

We excite the system with a uniform electric field pulse 
\begin{equation}
\vec{E}(t) = \vec{E}_0 \sin(\omega_0(t-t_0)) e^{-\frac{(t-t_0)^2}{2\tau^2}}
\end{equation}
centered at time $t_0$, with frequency $\omega_0$, peak amplitude $E_0$, and polarization direction $\hat{E}_0$. The Gaussian envelope with $\tau = 2.6$~fs corresponds to a full width at half maximum (FWHM) of $6.2$~fs. The effect of this pulse is to produce time-dependent hopping parameters (Peierls substitution \cite{aoki2014})
\begin{equation}
	h^{ij}_{\alpha\beta}(\mathbf{R},t) =  h^{ij}_{\alpha\beta}(\mathbf{R},t=0) e^{-\frac{ie}{\hbar} \phi_{ij}(\mathbf{R},t)},
\end{equation}
with the Peierls phase $\phi_{ij}(\mathbf{R},t)= - \int_0^t dt^\prime \vec{E}(t')\cdot(\vec{\mathbf{r}}_{j}-\vec{\mathbf{r}}_{i}+\mathbf{R})$. Here, $\vec{\mathbf{r}}_{i}$ is the position of site $i$. 

For the local interaction term, we choose the Kanamori form 
\begin{align}
\mathcal{H}_{\text{K}}^{i} =& \sum_{\alpha} U n^{i}_{\alpha \uparrow} n^{i}_{\alpha \downarrow} + \sum_{\alpha \neq \beta} U' n^{i}_{\alpha \uparrow} n^{i}_{\beta \downarrow}+ \sum_{\alpha<\beta, \sigma} (U'-J)n^{i}_{\alpha \sigma} n^{i}_{\beta \sigma}\nonumber\\
&-J \sum_{ \alpha \neq \beta} d_{\alpha \uparrow}^{i\dagger} d^{i}_{\alpha \downarrow}d_{\beta \downarrow}^{i\dagger} d^{i}_{\beta \uparrow}+J \sum_{\alpha \neq \beta} d_{\alpha \uparrow}^{i\dagger} d^{i\dagger}_{\alpha \downarrow}d^{i}_{\beta \downarrow} d^{i}_{\beta \uparrow}, 
\end{align}
where $U$ is the on-site intra-orbital Coulomb repulsion, $U^\prime$ the on-site interaction between different orbitals $\alpha$ and $\beta$, and $J$ the Hund coupling. We estimate the interaction parameters using the constrained random-phase approximation (cRPA) \cite{aryasetiawan2004}. The static values obtained with the RESPACK \cite{nakamura2021} code are $U=2$~eV and $J=0.3$~eV, which is substantially smaller than the $U=4$-$5$~eV and $J=0.7$-$0.9$ eV used in previous DMFT studies \cite{held2001,keller2004,poteryaev2007,toschi2010,hansmann2013}. 
These static $U$ values neglect the frequency dependence and hence underestimate the correlation effects. We therefore increase the intra-orbital interaction to $U=2.8$~eV to obtain an insulator with a gap size of approximately 0.3 eV, consistent with experiments~\cite{lupi2010,babich2020}.

We employ nonequilibrium DMFT~\cite{eckstein2013,aoki2014} to compute the evolution of the laser-excited lattice system. 
In our real-space DMFT calculations, we separately solve four impurity models corresponding to the four atoms within one unit cell.
As in previous materials studies, we use a simplified self-consistency \cite{petocchi2019,petocchi2023, chen2023}, which allows us to circumvent the explicit calculation of the lattice self-energies and lattice Green's functions, and we employ a noncrossing approximation (NCA) impurity solver~\cite{keiter1971,eckstein2010}. The initial temperature is $T=\frac{1}{12}~\text{eV}$. (Room temperature is too low for the NCA solver, but we do not expect qualitative differences due to the 0.3~eV gap.) To help with the interpretation of the spectra, we also solve the isolated dimer system using exact diagonalization (ED).

\begin{figure}[t]
	\includegraphics[width=0.99\linewidth]{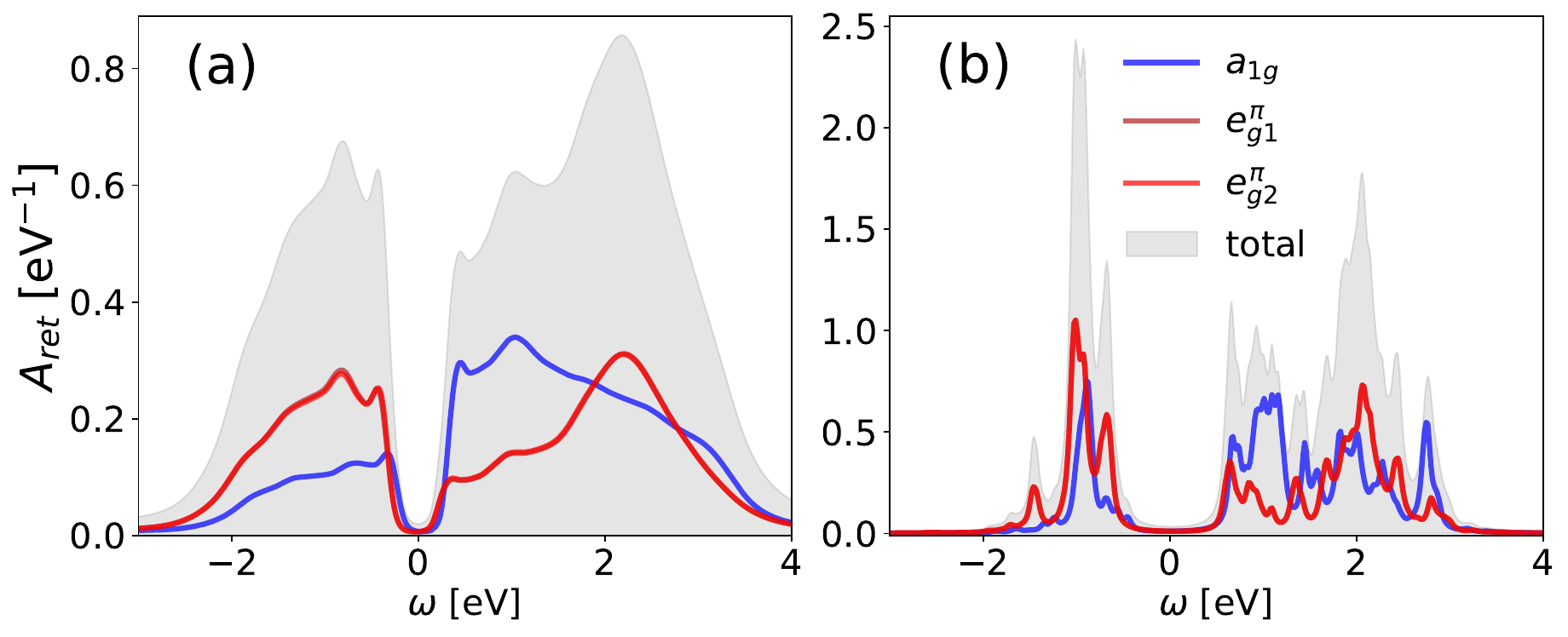}
	\caption{Orbital-resolved and total spectra in equilibrium obtained with DMFT (a) and ED (b). Both calculations yield a gap. 
	}
	\label{fig:spectra_eq}
\end{figure}

\section{Results}\label{sec:results}

\subsection{Equilibrium system}

We first discuss the equilibrium spectra obtained from DMFT and ED. As shown in Fig.~\ref{fig:spectra_eq}(a), the DMFT spectrum has a gap of 0.3 eV and the spectral weight below the Fermi energy is contributed by all three $t_{2g}$ orbitals. In the interacting system, the relative occupations of the orbitals are ($a_{1g} : e_{g1}^\pi : e_{g2}^\pi $) = (0.38 : 0.81 : 0.81), in good agreement with results of polarized X-ray spectroscopy experiments, which reported the ratios ($0.4 : 0.8 : 0.8$) \cite{park2000}. Similar results were also obtained in the previous LDA+DMFT studies~\cite{held2001,keller2004,poteryaev2007,toschi2010,rodolakis2010,hansmann2013}.

\begin{figure}[t]
	\includegraphics[width=0.8\linewidth]{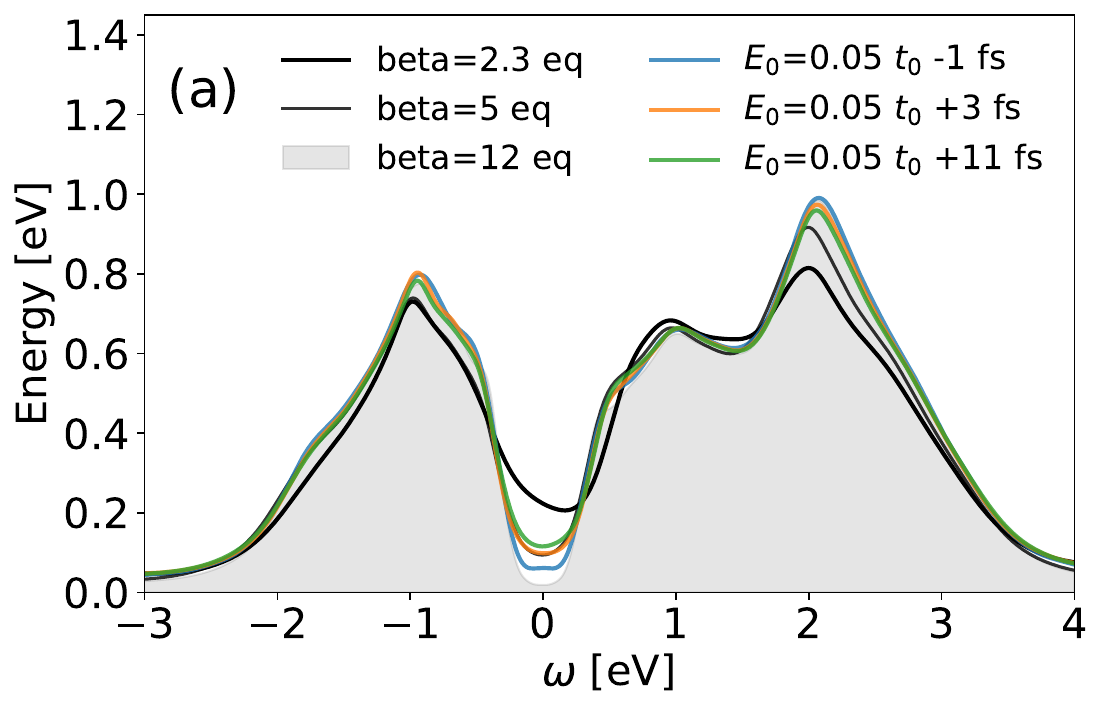}
	\includegraphics[width=0.8\linewidth]{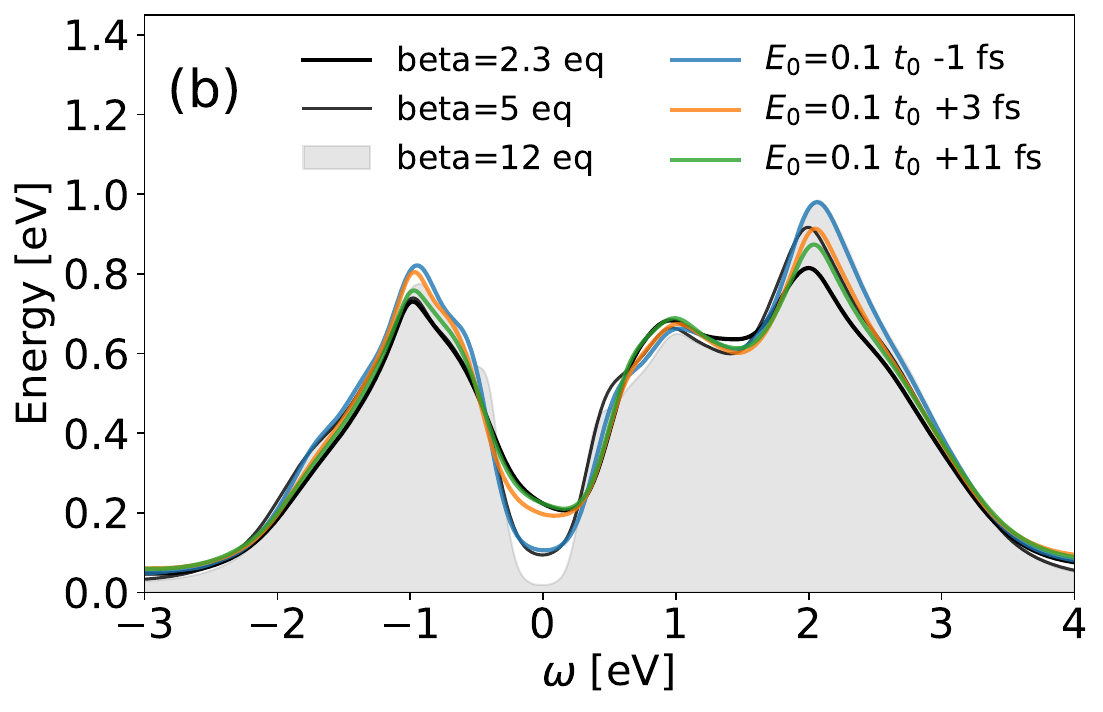}
	\caption{
	Total spectra for different temperatures (black or gray) and nonequilibrium spectra after the pulse excitation with $\omega_0=2.7$~eV (colored lines). Panel (a) is for pulse amplitude $E_0=0.05$~V/\textup{\AA} and panel (b) for $E_0=0.1$~V/\textup{\AA}.
	}
	\label{fig:spectra_ex}
\end{figure}

The ED spectrum of a single dimer with the same interaction and intra-dimer hopping parameters is shown in Fig.~\ref{fig:spectra_eq}(b). Based on the ED analysis, we identify a degenerate group of nine low-energy states (``ground states") with an approximate $S=1$ $e_{g1}^\pi e_{g2}^\pi$ configuration on each site. Here and in the following, ``$ab$ configuration" refers to a state with one electron in the $a$ orbital and one electron in the $b$ orbital, on the same site. Only $0.03$~eV higher in energy, there exists another group of nine states with mixed $e^\pi_{g1}e^\pi_{g2}$ and $e^\pi_ga_{1g}$ configuration. Even though at the LDA level, the trigonal spitting pushes the $a_{1g}$ bonding orbital below the $e_g^\pi$ orbitals, in the presence of interactions, the high-spin configurations with two electrons in the $e^\pi_g$ orbitals are preferred. 
The $0.03$~eV gap between the two groups of low energy states corresponds roughly to room temperature and explains the experimentally observed occupation of both the $e_{g1}^\pi e_{g2}^\pi$ and $e^\pi_ga_{1g}$ states \cite{park2000}.

The temperature dependence of the DMFT spectra is shown by the gray shading and black lines in Fig.~\ref{fig:spectra_ex}. Here, $\beta$ is the inverse temperature in eV$^{-1}$, so that $\beta=2.3$~eV$^{-1}$ corresponds to  $T\approx 5000$~K. We can see that an electronic temperature of several thousand K, as expected after a pump excitation, leads to a significant filling of the gap.

\begin{figure}[t]
	\centering
	\includegraphics[width=0.99\linewidth]{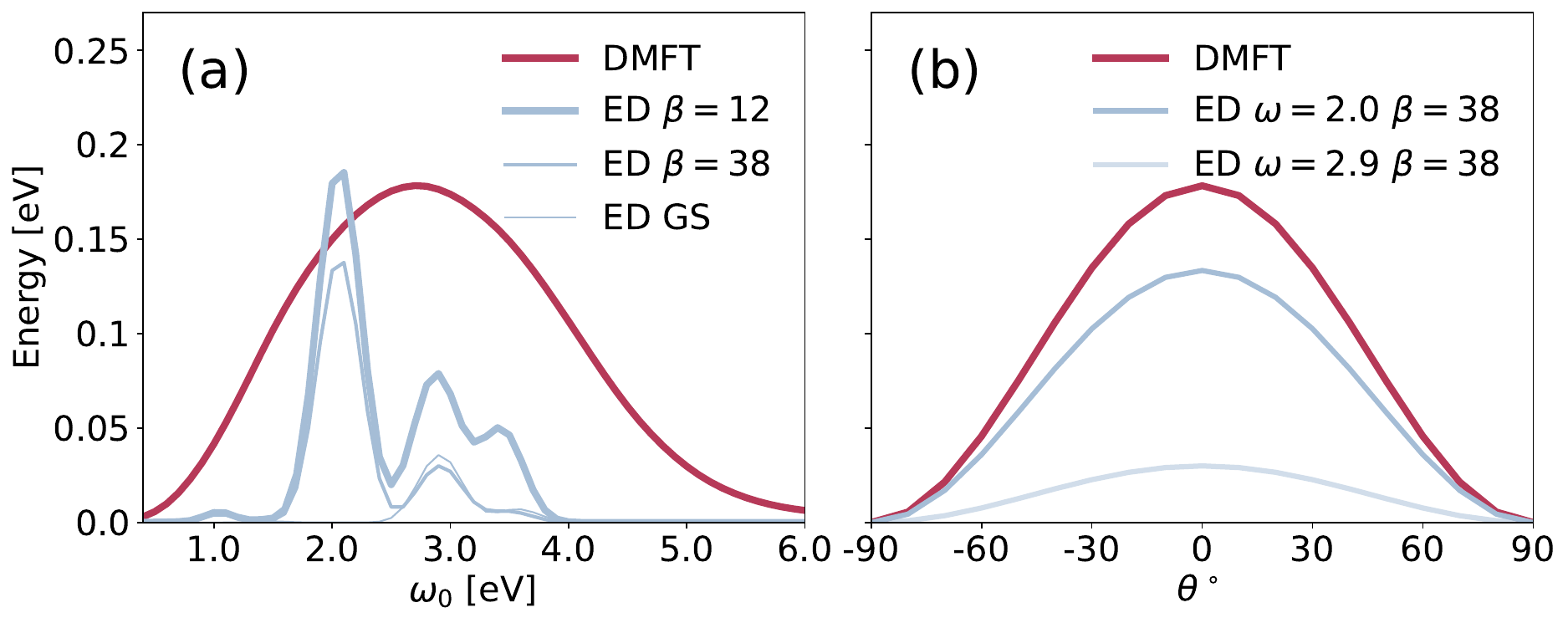}
	\caption{(a) Energy absorption as a function of the pulse frequency $\omega_0$ with $\theta=0^\circ$. (b) Energy absorption as a function of the polarization angle with frequency $\omega_0=2.7$~eV. $\theta=0^\circ$ corresponds to a polarization along the dimer, in the $\mathbf{c}_R$ direction.
	}
	\label{fig:absorption-frequency}
\end{figure}

\subsection{Pulse-excited system}

We first search for the polarization direction $\hat{E}_0=\vec{E_0}/|E_0|$ and pump frequency $\omega_0$ which yields the maximum energy absorption and then simulate the time evolution using nonequilibrium DMFT. The polarization angle $\theta$  between $\hat{E}_0$ and the dimerization axis ${\bf c}_R$, as well as the laser frequency $\omega_0$ are varied in the relevant range and the resulting energy absorption is shown in Fig.~\ref{fig:absorption-frequency}. Both in the DMFT and ED simulations, $\theta=0^\circ$ maximizes the absorption. With this polarization fixed, the DMFT simulations predict a broad energy absorption peak with a maximum corresponding to the pulse frequency $\omega_0=2.7$~eV, see red curve in Fig.~\ref{fig:absorption-frequency}(a). In the ED analysis of the dimer, the energy absorption spectra strongly depend on the temperature (see blue lines). As discussed before, the ground state mainly involves high-spin $e_{g1}^\pi e_{g2}^\pi$ configurations. Therefore, the laser excitation of the ground state creates doublons in the $e^\pi_g$ orbitals, i.~e., configurations where one of the $e^\pi_g$ orbitals is occupied by two electrons with spin up and spin down, 
\begin{equation}
\left(e_{g1}^{\pi\uparrow} e_{g2}^{\pi\uparrow}\right)_A+\left(e_{g1}^{\pi\downarrow} e_{g2}^{\pi\downarrow}\right)_B \rightarrow\left(e_{g1}^{\pi\uparrow}\right)_A+\left(e_{g1}^{\pi\downarrow} e_{g2}^{\pi\uparrow\downarrow}\right)_B \nonumber,
\end{equation}
 at the cost of $U=2.8$~eV (here A and B are the two atoms in a dimer). At high temperature, the thermal state includes both $e^\pi_{g1} e^\pi_{g2}$ and $e^\pi_g a_{1g}$ configurations, which activates additional excitations with energy $U-3J=1.9$~eV, given for example by
\begin{equation}
\left(e_{g1}^{\pi\uparrow}e_{g2}^{\pi\uparrow}\right)_A
+\left(e_{g1}^{\pi\uparrow} a_{1g}^{\uparrow}\right)_B \rightarrow\left(e_{g1}^{\pi\uparrow}\right)_A
+\left(e_{g1}^{\pi\uparrow} e_{g2}^{\pi\uparrow} a_{1g}^{\uparrow}\right)_B,\nonumber
\end{equation}
and with energy $U+2J=3.4$~eV, for example 
\begin{align}
\left(e_{g1}^{\pi\uparrow}e_{g2}^{\pi\uparrow}\right)_A
+\left(e_{g2}^{\pi\downarrow} a_{1g}^{\downarrow}\right)_B \rightarrow&\left(e_{g1}^{\pi\uparrow}\right)_A\nonumber\\
&+\left(\alpha e_{g1}^{\pi\downarrow} e_{g2}^{\pi\uparrow\downarrow}+ \gamma e_{g1}^{\pi\downarrow} a_{1g}^{\uparrow\downarrow}\right)_B, \nonumber
\end{align}
similar to the three-band Kanamori-Hubbard model in the atomic limit with degenerate orbitals \cite{de_medici_janus-faced_2011}. We conclude that the multi-orbital interactions play a crucial role in the photo-excitation process. The initial temperature of the system is also important, because it controls the initial state population and and influences the charge dynamics. 

\begin{figure*}[t]
	\centering
	\includegraphics[width=0.34\textwidth]{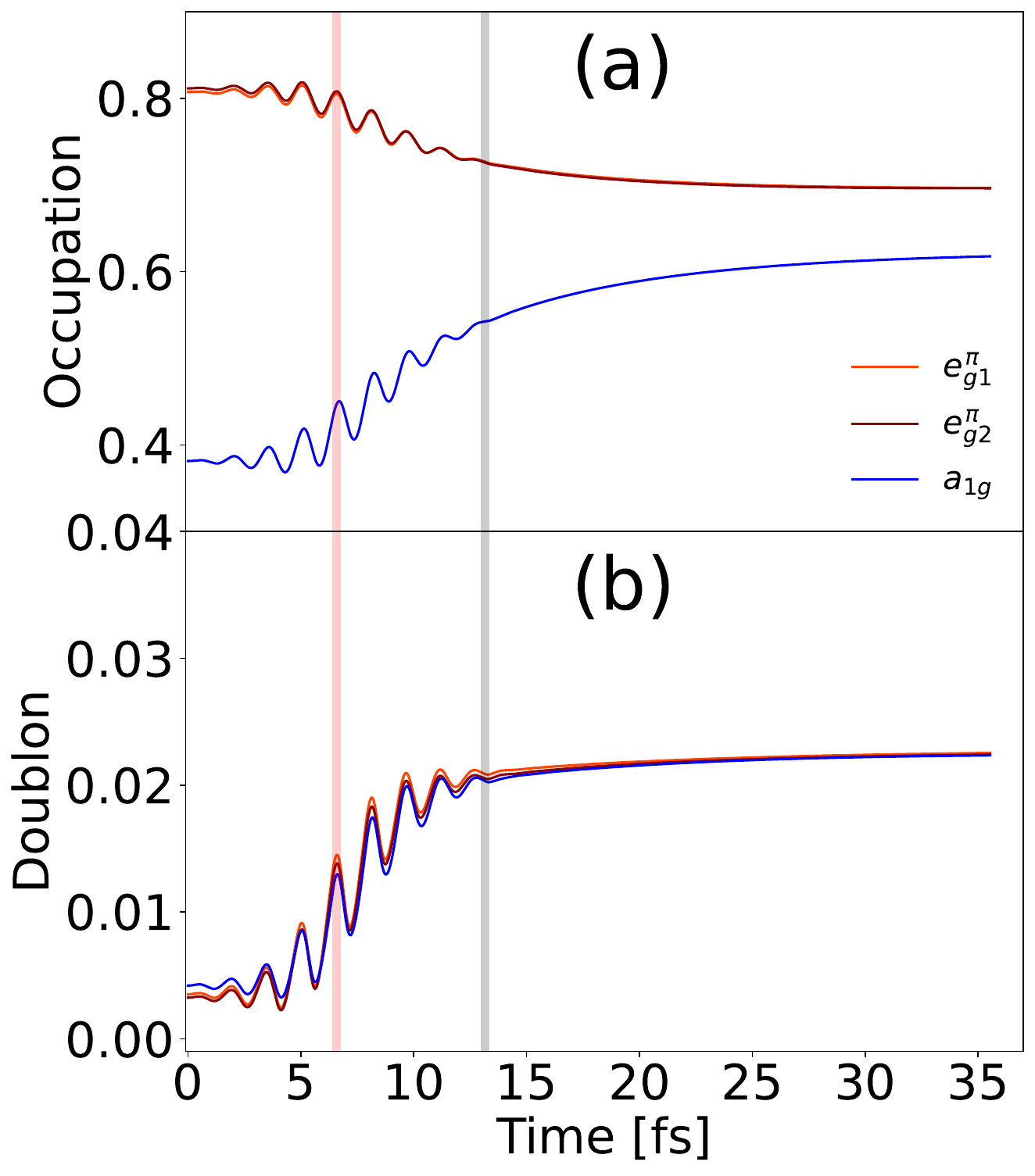}
	\includegraphics[width=0.6\textwidth]{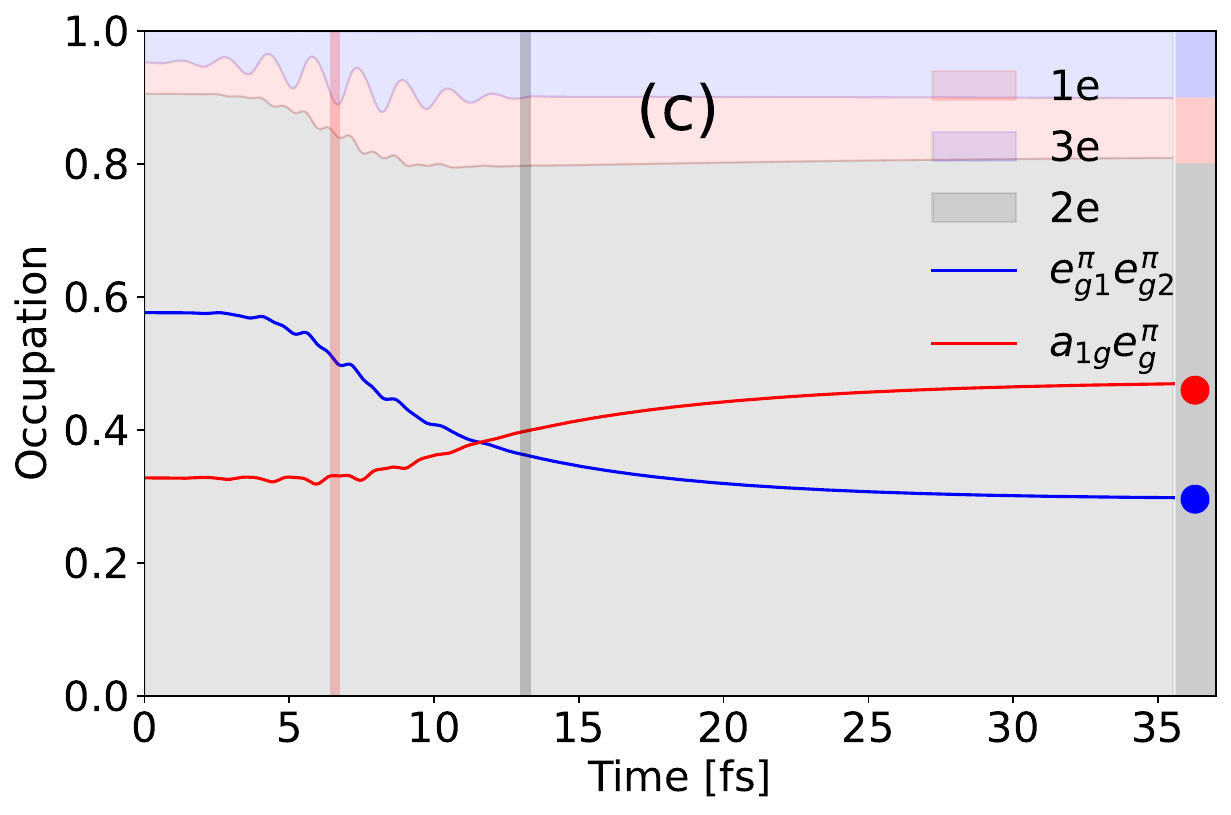}
	\caption{ Occupation (a) and double occupation (or doublon density) (b) in each of the three orbitals, for a given vanadium atom. 
	(c) Projection of the state occupation onto electron number sectors (colored shadings), for a given vanadium atom. Within the $n=2$ sector, the states are further projected onto the indicated orbital configurations. The red and gray vertical lines indicate the maximum and the end of the laser pulse, respectively. The colored bars and dots at the right side of the panel indicate the weights of the charge sectors and the weights of the $e^{\pi}_{g1}e^{\pi}_{g2}$ and $e^{\pi}_ga_{1g}$ configurations in the thermalized state (equilibrium system with the same energy as the system after the excitation).			 
   }
	\label{fig:occupation}
\end{figure*}

The DMFT results for the time-dependent orbital occupations are shown in Fig.~\ref{fig:occupation}(a). In this study, we perform single-site DMFT calculations (in contrast to the cluster DMFT calculations of Ref.~\cite{chen2023}) and report the population dynamics for a single V atom. 
During the pulse, which reaches its maximum at $t=t_0=6.6$~fs (indicated by the red vertical line), the $t_{2g}$ electrons are driven back and forth between the vanadium atoms, as can be deduced from the oscillations of the occupations on a given site. This charge sloshing, and other effects of the laser pulse like bandwidth renormalization \cite{dunlap1986,tsuji2011}, also lead to a modulation in the double occupation, as seen by the oscillating doublon density in Fig.~\ref{fig:occupation}(b). We note that the DMFT simulation represents an infinite bulk system and captures interaction and heating effects, i.e., the conversion of the injected energy into various electronic excitations. Figure~\ref{fig:occupation} reveals a significant rearrangement of charge between the different orbitals. In particular, panel (a) shows a significant flow of charge from the  $e^\pi_g$ to the  $a_{1g}$ orbitals during the pulse, which ends approximately at $\sim$13~fs (grey vertical line), followed by a slower reshuffling and saturation around 35 fs. The average doublon density increases during the pulse, but remains low. 

 As illustrated in Fig.~\ref{fig:occupation}(c), in the equilibrium PI phase, the sector with $n=2$ electrons (gray shading) represents more than 90\% of the weight of the initial state. The laser pulse mainly reshuffles the states within this $n=2$ sector, but also produces some states with $n=1$ and $n=3$ through charge excitations between vanadium atoms. These charge excitations are indicated by the red and violet shading. After the end of the pulse ($t\gtrsim 13$~fs), the weight of the $n=2$ sector remains almost unchanged, and in our closed system (without energy dissipation to phonons or other external degrees of freedom), the electrons are expected to thermalize at $T\approx 5000$~K. This is the temperature at which the energy of the equilibrium system equals the energy of the simulated system after the pulse. The corresponding weights of the charge sectors and the weights of the $e^{\pi}_{g1}e^{\pi}_{g2}$ and $e^{\pi}_ga_{1g}$ configurations are indicated by the colored bars and dots at the right side of panel (c).

Within the two-electron sector, the $e^{\pi}_{g1}e^{\pi}_{g2}$ configurations of the ground state manifold dominate the PI phase in equilibrium. During and after the pulse, the fidelity of these configurations decreases rapidly, while the  $a_{1g}e^{\pi}_g$ fidelity increases rapidly, consistent with panel (a). After the pulse, the $a_{1g}e^{\pi}_g$ configurations become dominant. An important finding is that the weights of the different charge sectors and configurations are very close to the thermal reference values already 20 fs after the pulse, and that there is no indication of a trapping in a nonthermal transient state. This is in contrast to the finding for the related material VO$_2$ in Ref.~\cite{chen2023}, where due to the V$^{4+}$ configuration, there is one electron in the three $t_{2g}$  orbitals and where the much stronger dimerization locks this electron in the lowest lying bonding orbital, well separated from other states. 

The population of the $a_{1g} e^{\pi}_g$ states generates spectral weight in the gap region and results in the rapid partial gap filling seen in Fig.~\ref{fig:spectra_ex}(b). Our calculations thus demonstrate that the metallic phase observed after photo-doping is a consequence of the charge reshuffling between $e^{\pi}_g$ and $a_{1g}$ orbitals. But also the nonequilibrium spectrum after the pulse is very similar to a thermal spectrum corresponding to a high electronic temperature ($\beta \approx 2.3 \text{ eV}^{-1}\leftrightarrow T\approx 5000$~K), see the comparison to the thermal results (black gray curves). 

\begin{figure*}[t]
	\centering
	\includegraphics[width=0.32\linewidth]{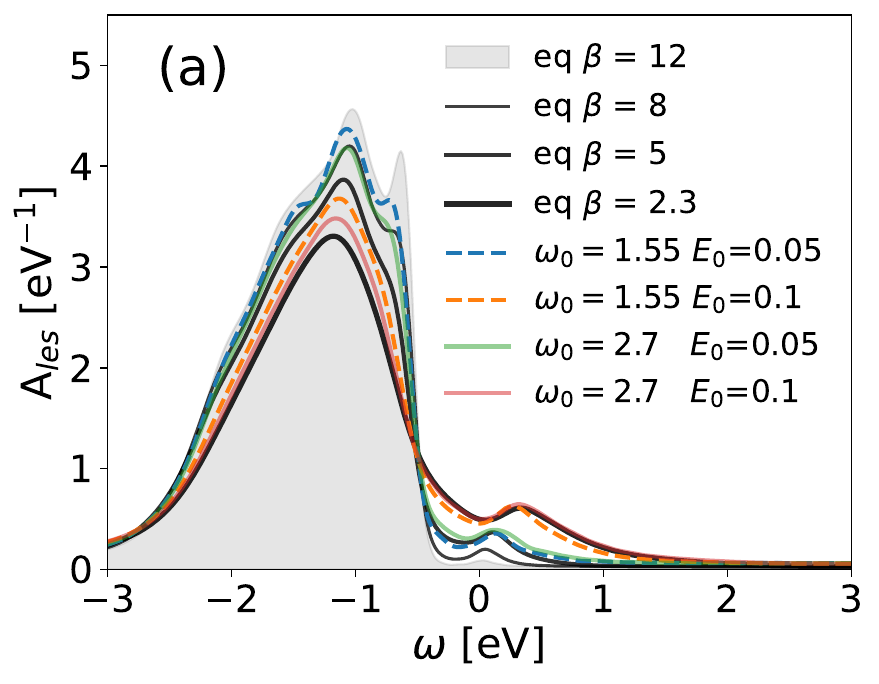}
	\includegraphics[width=0.32\linewidth]{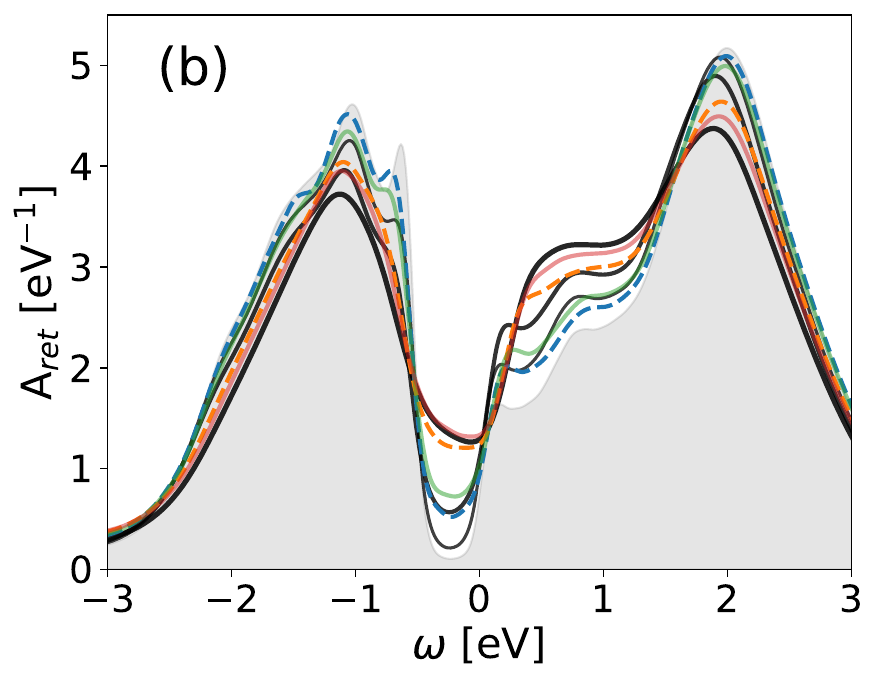}
	\includegraphics[width=0.32\linewidth]{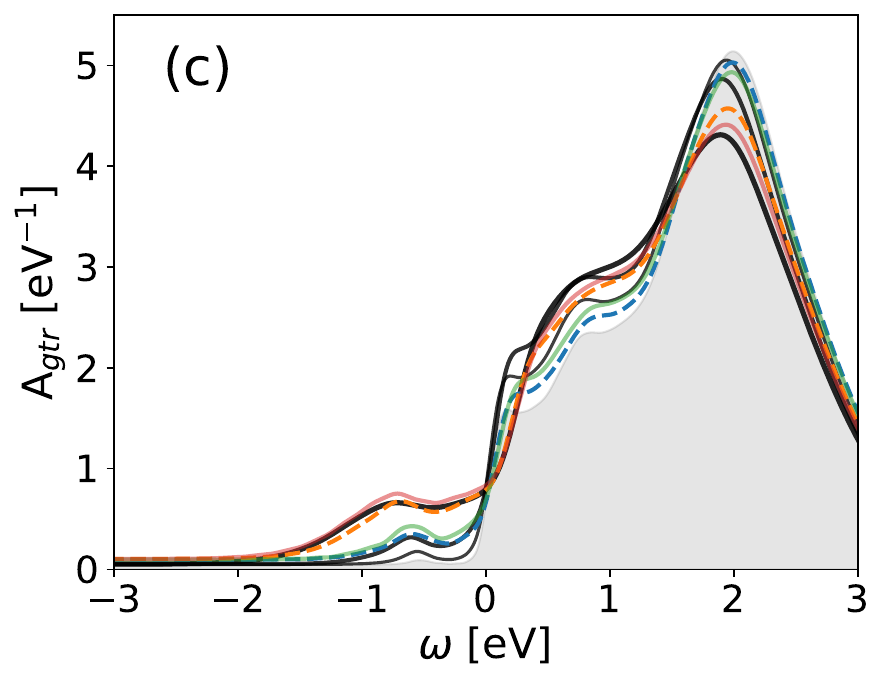}
	\caption{
	Local lesser (a), retarded (b) and greater (c) spectra obtained with DMFT in equilibrium for different temperatures (gray shading and black lines) 
	and 17~fs after the pulse excitation for the indicated $E_0$ [V/\textup{\AA}] and $\omega_0$~[eV]. The initial inverse temperature is $\beta=12$~eV$^{-1}$.
	}
	\label{fig:ppsc_spectra}
\end{figure*}

In Fig.~\ref{fig:ppsc_spectra}, we present the time-dependent DMFT spectra, obtained for the optimal polarization ($\theta=0^\circ$) and absorption frequency ($\omega_0=2.7$~eV) as well as the typical laser frequency in photo-doping experiments~\cite{lantz2017,babich2020} ($\omega_0=1.55$~eV). The total spectral functions $A^{\text{ret}}$, electron occupation functions $A^<$ and hole occupation functions $A^>$  are shown for probe time $t_0=17$~fs and two pulse amplitudes. These spectra are obtained from the imaginary parts of the corresponding energy-dependent Green's functions, which we calculate using the Wigner transformation~\cite{aoki2014} 
\begin{equation}
	G^X\left(\omega, t_{\mathrm{av}}\right)=\int d t_{\mathrm{rel}} e^{i \omega t_{\mathrm{rel}}} G^X\left(t, t^{\prime}\right). 
\end{equation}
Here, $t_{\mathrm{rel}} \equiv t-t^{\prime}$ is the relative time, $t_{\mathrm{av}}\equiv t+t^{\prime}$ is the averaged time, and $X$ refers to the retarded, lesser and greater components of the contour Green's function. $A^<(\omega,t_\text{av})$ approximately represents the time-resolved photoemission spectrum. For both pulse frequencies, the laser excitation results in a partial filling of the gap, which is slightly more pronounced for the higher energy excitation (since the larger amount of injected energy results in stronger heating).  

\begin{figure}[t]
	\centering
	\includegraphics[width=0.9\linewidth]{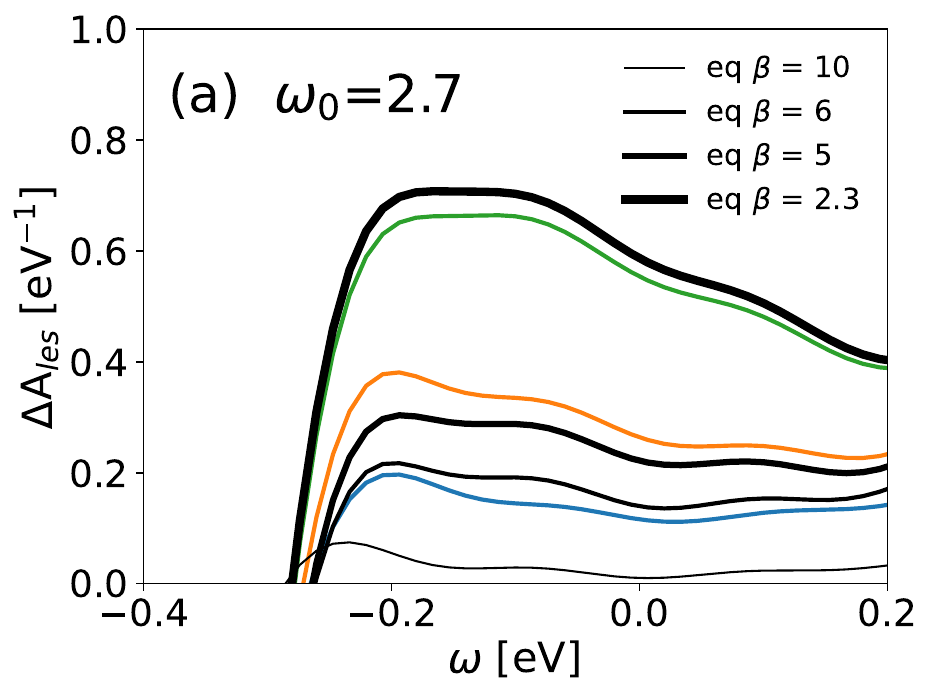}
	\includegraphics[width=0.9\linewidth]{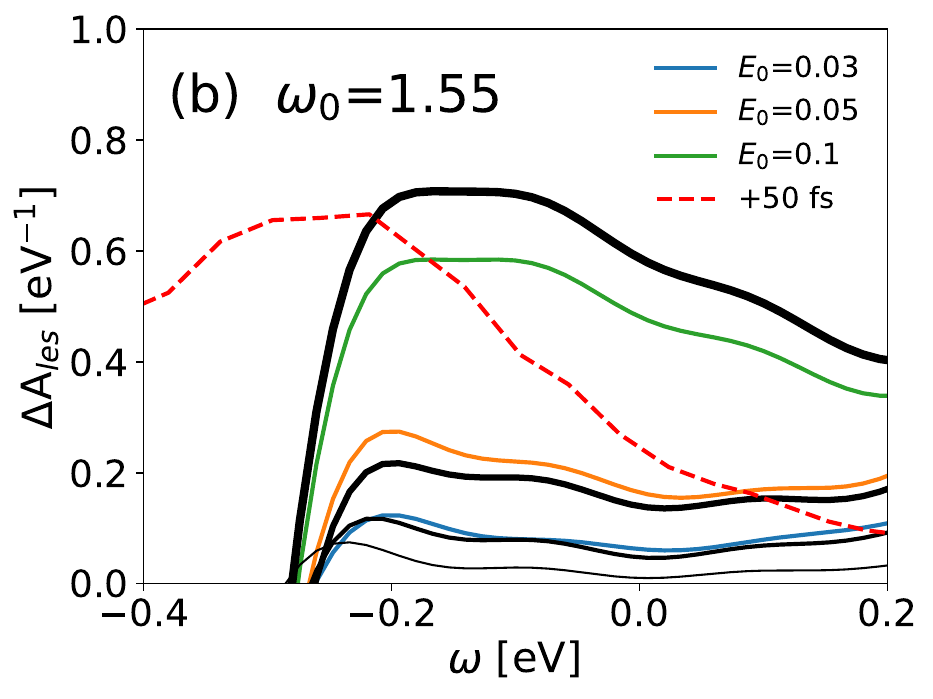}
	\caption{
	Difference in the occupation from the equilibrium value for $\beta=12$~eV$^{-1}$. The nonequilibrium distributions are measured at time $t=17$~fs for a pulse with $\omega_0=2.7$~eV (panel (a)) and 1.55~eV (panel (b)).  In panel (b), the dashed line shows experimental data extracted from Fig.~2(b) of Ref.~\cite{lantz2017} (with an arbitrary rescaling factor) for comparison.
	}
	\label{fig:diff_spectra}
\end{figure}

In Fig.~\ref{fig:diff_spectra} we plot the difference between the nonequilibrium occupation at $t=17$~fs and the initial equilibrium occupation for $\beta=12$~eV$^{-1}$ (green, orange and blue lines), as well as differences between higher-temperature equilibrium occupations and the initial occupation (black lines). Panel (a) is for the $\omega_0=2.7$~eV pulse and panel (b) for the $\omega_0=1.55$~eV pulse. Consistent with the previous observations, we find that the nonequilibrium occupation at $t=17$~fs is almost thermalized (see data for $\omega_0=2.7$~eV, where the thermalized system has $\beta=2.3$~eV$^{-1}$). Comparing our results to Fig.~2(b) of Ref.~\cite{lantz2017}, we see that their difference $\Delta I$ in photoemission spectra between the pumped and unpumped systems looks qualitatively similar for the shortest reported delay time of 50~fs \footnote{At higher energies, around $\approx 0.3$~eV the theoretical occupations show an upturn associated with the upper Hubbard band, which is not evident in the measured spectra}. We note that since the initial system is insulating, the location of the Fermi energy could be somewhat different in the experiment, which may explain the horizontal shift between the theoretical and experimental curves. 

On a timescale of several hundred fs, $\Delta I$ vanishes inside and above the gap region, while after two ps, it approaches a distribution consistent with a slightly increased equilibrium temperature. In view of our simulation data, which show no evidence for a delayed thermalization in the electronic system, these experimental results are most naturally interpreted in terms of a two temperature picture \cite{allen1987,singh2010}, where the coupling of the hot electrons to the lattice results in a cooling of the electronic distribution on a ps timescale.

\section{Conclusions}\label{sec:conclusions}

Our  {\it ab initio} nonequilibrium DMFT simulations of Cr-doped V$_2$O$_3$ clarified the photoinduced charge dynamics during and after a laser pulse applied to the PI phase. The optical excitation induces an ultrafast gap filling by generating inter-orbital electron pairs, which thermalize within tens of fs to a high-temperature metallic state. On a longer timescale, the lattice is expected to cool down the electronic system, and an explicit simulation of this energy reshuffling would require the introduction of phonon degrees of freedom. For the short-time dynamics, our study shows that multi-orbital Mott and Hund physics play a key role in the formation of dominant $S=1$ atomic configurations in equilibrium, and in determining the energy absorption from an optical pulse.

While the experimentally measured photoinduced metal states may be nonthermal in the sense of different electronic and lattice temperatures, our analysis does not provide any clear indication for the existence of a long-lived photo-induced nonequilibrium metallic state of the electronic subsystem in Cr-doped V$_2$O$_3$. This is in contrast to the situation found in photo-doped VO$_2$~\cite{chen2023}, where the simpler and sparser energy level structure leads to distinctly nonthermal metal states that last for hundreds of fs. 

The simulations presented in this work assumed a homogeneous system, with the Cr doping merely affecting the lattice structure. The possible role of inhomogeneities on the excitation and relaxation pathway in (V$_{1-x}$Cr$_{x}$)$_2$O$_3$ ($x=0.038$) is an interesting topic for future studies.
\acknowledgements
This work was supported by the Swiss 
National Science Foundation via the Research Unit QUAST 
of Deutsche Foschungsgemeinschaft (FOR5249) (J.C.). 
The calculations were run on the beo05 
cluster at the University of Fribourg, using a code based on NESSi \cite{schuler2020}.

\bibliography{ref}

\end{document}